\documentclass[ reprint,  amsmath,amssymb,  aps, prb]{revtex4-1}

\usepackage{graphicx}
\usepackage{dcolumn}
\usepackage{bm}

\begin{document}

\preprint{APS/123-QED}
\title{Stability of multielectron bubbles in high Landau levels}


\renewcommand{\thefootnote}{\fnsymbol{footnote}}

\author{Dohyung Ro$^{1}$, S.A. Myers$^{1}$, N. Deng$^{1}$, 
J.D. Watson$^{1}$, M.J. Manfra$^{1,2,3}$, L.N Pfeiffer$^5$, 
K.W. West$^5$, and G.A. Cs\'{a}thy$^{1,3}$}

\affiliation{
${}^1$Department of Physics and Astronomy, Purdue University, West Lafayette, IN 47907, USA \\
${}^2$School of Materials Engineering and School of Electrical and Computer Engineering, 
          Purdue University, West Lafayette, IN 47907, USA \\
${}^3$Birck Nanotechnology Center Purdue University, West Lafayette, IN 47907, USA \\
${}^4$Department of Electrical Engineering, Princeton University, Princeton, NJ 08544, USA 
}

\date{\today}

\begin{abstract}

We study multielectron bubble phases in the $N=2$ and $N=3$ Landau levels in a high mobility GaAs/AlGaAs sample.
We found that the longitudinal magnetoresistance versus temperature curves
in the multielectron bubble region exhibit sharp peaks, irrespective of the Landau level index. 
We associate these peaks with an enhanced scattering caused by thermally fluctuating 
domains of a bubble phase and a uniform uncorrelated electron liquid at the onset of the bubble phases. 
Within the $N=3$ Landau level,
onset temperatures of three-electron and two-electron bubbles exhibit linear trends with respect to the filling factor;
the onset temperatures of three-electron bubbles are systematically higher than those of two-electron bubbles.
Furthermore, onset temperatures of the two-electron bubble phases across $N=2$ and $N=3$ Landau levels
are similar, but exhibit an offset. This offset and the dominant nature of the three-electron bubbles in
the $N=3$ Landau level reveals the role of the short-range part of the electron-electron interaction 
in the formation of the bubbles.
%These measurements offer information on bubble energetics that is expected to lead to improvements
%of existing theories and that reveals the impact of the short-range part of the effective electron-electron 
%interaction driving bubble formation.

\end{abstract}

%\pacs{Valid PACS appear here}
\maketitle

The two-dimensional electron gas (2DEG) exposed to perpendicular magnetic fields is a rich model system 
that hosts a variety of electronic phases. Perhaps the most well known of these phases are
the fractional quantum Hall states\cite{tsui} which harbor topological order.
Electron solids possessing charge order form yet another distinct group of phases. 
Examples of electronic solids are the Wigner solid\cite{ws0}, electronic bubble phases, and
quantum Hall nematic or stripe phases\cite{fogler,fogler96,moessner,lilly,du,cooper,jim2,jim-rev,nem1,nem2}.

Bubble phases are among the most recently discovered phases of 2DEGs
which have not yet revealed all their properties. They were
predicted by a Hartree-Fock theory\cite{fogler,fogler96,moessner} and confirmed by exact diagonalization\cite{haldane}
and density matrix renormalization group  studies \cite{shibata01} to be a periodic arrangement of clusters or bubbles of electrons.
In linear transport bubble phases are identified by reentrant integer quantum Hall behavior\cite{du,cooper,jim2}.
In addition microwave absorption\cite{lewis,lewis3,lewis4}, non-linear transport\cite{cooper,nl2,nl3,nl4,nl5,nl6}, 
surface acoustic wave propagation\cite{msall,smet}, temperature 
dependence\cite{xia,pan,nuebler,deng1,deng2,deng3}, and thermopower
measurements\cite{chick,chick-1} also support the formation of bubbles.
However, we still lack direct probes of the morphology of the bubbles. 

Bubble phases are commonly observed in 2DEGs in GaAs/AlGaAs\cite{lilly,du,jim-rev,jim2,
lewis,lewis3,lewis4,cooper,nl2,nl3,nl4,nl5,nl6,msall,smet,xia,pan,nuebler,deng1,deng2,deng3,chick,chick-1}
and have also been recently
seen in graphene\cite{dean}. In the former system, bubbles form in high Landau levels,
at orbital Landau level index $N$ greater or equal to 1. Here we used the customary labeling of
quantum numbers of energy levels associated with cyclotron motion, $N=0$ being the lowest Landau level. 
Theories allow for different types of bubble phases within a given Landau 
level\cite{fogler97,foglerBook,cote03,goerbig03,yoshi03,goerbig04,yoshi04,dorsey06,yoshi02}.
The different types of bubble phases are distinguished by the number of electrons per
bubble $M$; a modest change in the Landau level filling factor was predicted to result in a phase
transition between two different types of bubble phases. 
Measurements for nearly two decades did not resolve such distinct bubble phases.
Only recently were distinct bubble phases observed in the $N=3$ Landau level\cite{zud,kevin}; 
the Landau level filling factors of these bubble phases were in 
excellent agreement with calculations. These observations allowed the assignment of the number 
of electrons per bubble for each bubble phase
and cemented the bubble interpretation of the reentrant integer quantum Hall states. 

\begin{figure*}[t]
  \centering
  \includegraphics[width=6.2in]{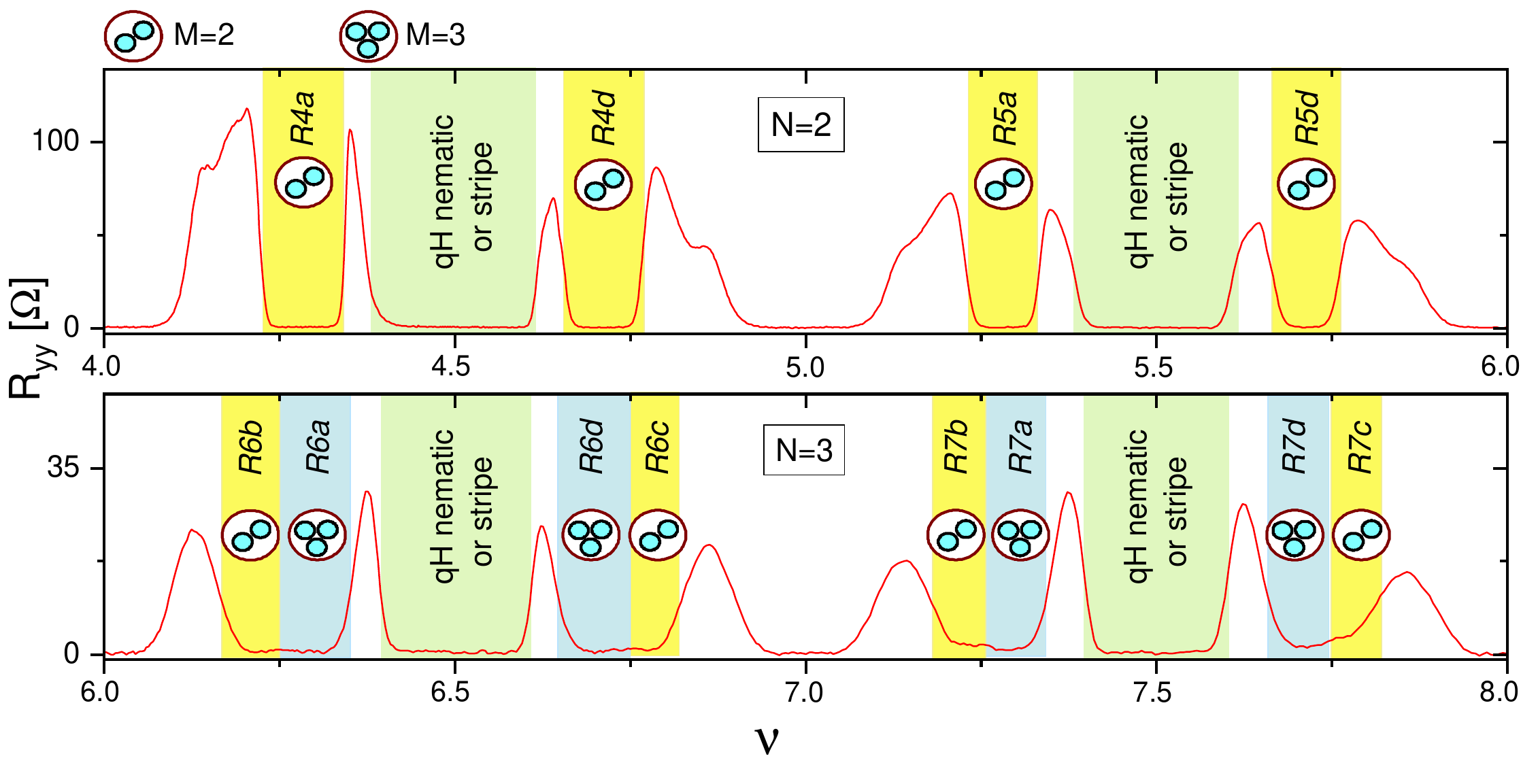}
  \caption{The dependence of the longitudinal magnetoresistance $R_{yy}$ on filling factor $\nu$
  in the $N=2$ (top panel) and $N=3$ (bottom panel) Landau levels. The two-electron bubble phases 
  ($M=2$) are shaded in yellow, whereas the three-electron bubble phases ($M=3$) are shaded in blue. 
  Vanishing $R_{yy}$ near integer filling factors indicate integer quantum Hall states, whereas areas shaded 
  in green near half-integer  filling factors are quantum Hall nematics. Data collected at $T=59$~mK.
  }
\end{figure*}

Recent observations of distinct multielectron
bubble phases {\it within} one Landau level\cite{zud,kevin}, that at $N=3$, 
opened up the possibility for their qualitative and quantitative analysis
both within one Landau level and also across different Landau levels.
We found that, in our high mobility GaAs/AlGaAs sample bubble phases in the $N=3$ Landau level
exhibit sharp peaks in the longitudinal magnetoresistance versus temperature curves,
as measured at fixed magnetic fields. Similar peaks were detected in the $N=1$ and $N=2$
Landau levels in high mobility GaAs/AlGaAs\cite{deng1,deng2,chick-1} and also in a  graphene sample\cite{dean},
but such peaks appear to be absent in a low mobility GaAs sample
containing alloy disorder\cite{zud}. We think these peaks are due to scattering through the bulk of the sample
when the bulk consists of interpenetrating and fluctuating domains of a bubble phase and a uniform uncorrelated liquid.
Within this interpretation, the temperature of the peak 
is identified with the onset temperature of the bubble phase. We found that the onset temperatures of the bubble phases 
determined this way have a linear trend with the filling factor and a particular dependence on the number of electrons
per bubble. Within the $N=3$ Landau level,
the onset temperatures of $M=3$ bubbles are higher than those of $M=2$ bubbles
and exhibit different trends with the filling factor.
Furthermore, when comparing the $M=2$ bubble phases across $N=2$ and $N=3$ Landau levels,
we find that onset temperatures are similar but exhibit an offset.
These measurements offer information on bubble energetics that may be used for a qualitative comparison to theories
and reveal details of the short-range part of the effective electron-electron interaction.

We measured a 2DEG confined to a $30$~nm wide GaAs/AlGaAs quantum well. This sample
has an electron density 
$n= 2.8 \times 10^{11}$~cm$^{-2}$ and mobility $\mu = 15 \times 10^{6}$~cm$^{2}$/Vs
and it is the same as the one reported on in Ref.\cite{kevin}.
In order to stabilize the temperature of the sample, we took advantage of the large heat capacity of liquid He-3
by mounting our sample in a He-3 immersion cell\cite{setup}. 
The temperature in this experiment is measured by a common resistive ruthenium oxide thermometer.
The sample is grown on the (100) face of GaAs
and it is cleaved into a $4 \times 4$~mm$^2$ square shape.

 \begin{figure}[t]
  \centering
  \includegraphics[width=3.3in]{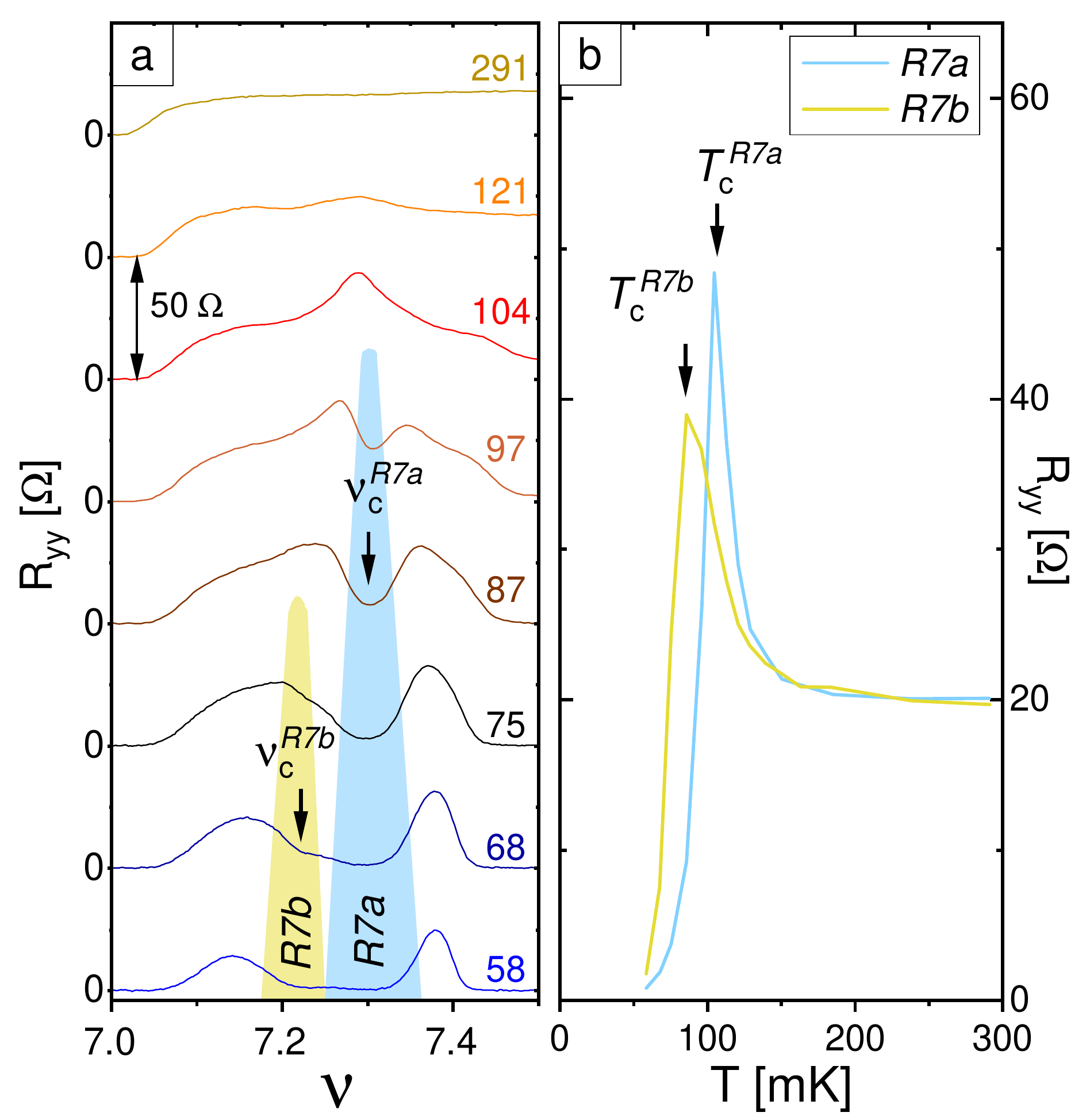} 
  \caption{(a) Isothermal evolution of $R_{yy}$ versus $B$-field for the $R7a$ and $R7b$ 
  bubble phases. Numbers on each trace are the temperature in units of mK.
  Arrows mark the central filling factors $\nu_c$ of these phases.
  (b) Evolution of $R_{yy}$ versus $T$ of the $R7a$ and $R7b$ 
  bubble phases at their respective central filling factors $\nu_c^{R7a}=7.30$, and $\nu_c^{R7b}=7.22$. 
  Arrows indicate the onset temperatures $T_c$ of these phases near the peak
  region of the  $R_{yy}$ versus $T$ curves.}
\end{figure}
 
In Fig.1 we show magnetotransport against the Landau level filling factor $\nu$
in the $N=2$ and $N=3$ Landau level. Here $\nu=h n/ e B$, $h$ is the Planck constant,
$e$ the elementary charge, and $B$ the applied magnetic field. Regions of vanishing
longitudinal resistance $R_{yy}$ in this figure are associated with a variety of phases. 
At integer values of the Landau level filling factor $\nu=i$, $R_{yy}=0$ and the Hall resistance
is quantized to $h/ie^2$, indicating integer quantum Hall states\cite{klitz}. Here $i=4,5,6,$ and $7$.
At half integer values $\nu=i+1/2$ there are quantum Hall nematics or stripes\cite{lilly,du}.
Finally, at other non-integer values of $\nu$ bubble phases form. In the $N=2$ Landau level
only one type of multielectron bubble phase develops\cite{du,cooper}. 
We extensively report on $R_{yy}$, the longitudinal 
magnetoresistance along the [110] crystallographic axis of our sample.
In the region of bubble phases, the magnetoresistance is nearly isotropic \cite{du,cooper,jim2,zud,kevin}.

In contrast to the $N=2$ Landau level, as recently discovered, in the $N=3$ Landau level
there are two different types of multielectron bubble phases\cite{zud,kevin}. 
Based on an excellent agreement 
of the measured filling factors of these phases with those predicted by the theory, 
the number of electrons per bubble was identified for each bubble phase. 
In Fig.1 we shaded and labeled the two-electron ($M=2$)
and three-electron ($M=3$) bubble phases. Multielectron bubble phases of the $N=3$ Landau level 
are separated by a small magnetoresistive feature\cite{zud,kevin}. The Hall resistance of bubble phases was found
to be quantized to integer values of the nearest integer quantum Hall plateau\cite{du,cooper,jim2,zud,kevin}
(not shown in Fig.1).
Using techniques other than transport, in these Landau levels $M=1$ bubbles also form as part of 
plateaus of integer quantum Hall states\cite{ws1,ws11,ws2,ws3,ws33,ws4}.
However, our transport experiments cannot distinguish them from other localized states
and thus in this Article the $M=1$ bubble phases will not be further discussed.

Similarly shaded bubble phases in Fig.1 appear to form at particle-hole conjugated filling factors\cite{deng2,zud,kevin}.
In the following we examine this apparent symmetry to a greater detail. Bubble phases in high mobility
samples, such as ours, form in a range of filling factors. We define $\nu_c$, 
the central filling factor of a bubble phase, as the filling factor of its highest stability. 
Thus the central filling factor is the filling factor of the local minimum in the longitudinal  magnetoresistance
in the bubble phase region that may confidently be detected at the highest possible temperature\cite{deng1,deng2}.
For example, in Fig.2a we observe that for the $R7a$ bubble phase 
there is a local minimum in $R_{yy}$ isotherms that persists to temperatures as high as $T=97$~mK. 
This local minimum is observed at $\nu_c^{R7a}=7.30$.
Similarly, for the more fragile $R7b$ phase there is a
local resistance minimum developed at $\nu_c^{R7b}=7.22$
at temperatures as high as $T=75$~mK.
These and central filling factors of other multielectron
 bubble phases of the $N=3$ Landau level are shown in Table.I. Errors for filling factors are $\pm 0.01$.
We notice that the central filling factors of the family of $M=3$ bubble phases
can be written in the form $\nu_c=6+0.30, 7-0.30, 7+0.30, 8-0.30$ for
$R6a$, $R6d$, $R7a$, and $R7d$, respectively. Furthermore, 
the filling factors of the family of $M=2$ bubble phases
can be written in the form  $\nu_c=6+0.22, 7-0.23, 7+0.22, 8-0.22$ for
$R6b$, $R6c$, $R7b$, and $R7c$, respectively.
We thus found that, similarly to the bubble phases of
the $N=1$ and $2$ Landau levels\cite{deng1,deng2}, those of the $N=3$ Landau level also form
at central filling factors related by particle-hole conjugation\cite{zud}.

\begin{table}[b]
 \caption{Central filling factors $\nu_c$ and onset temperatures $T_c$ of the bubble phases
 of the $N=3$ Landau level.}
 \begin{ruledtabular}
 \begin{tabular}{l c c c c c c c c}
                        & $R6a$ & $R6b$ & $R6c$ & $R6d$ & $R7a$ & $R7b$ & $R7c$ & $R7d$ \\
 \hline
 $\nu_c$               & 6.30 & 6.22 & 6.77 & 6.70 & 7.30 & 7.22 & 7.78 & 7.70 \\
 $T_c$[mK]               & 117  & 100  & 91  & 117  & 101  & 80  & 71   & 100 \\
 \end{tabular}
 \end{ruledtabular}
\end{table}

While the isotherm at $T=97$~mK in Fig.2a exhibits a local minimum near $\nu_c^{R7a}=7.30$,
that at  $T=104$~mK exhibits a local maximum. 
We define  $T_{c}^{R7a}$, the onset temperature of $R7a$, as the average of 
highest temperature at which there is a local minimum in $R_{yy}$ and the next highest temperature of measurement.
The difference between these two temperatures signifies the error in determining $T_{c}$.
Values obtained from such an analysis of this and other multielectron bubble phases of the $N=3$ Landau level
are found in Table.I. Errors for $T_{c}$ are $\pm 5$~mK.
We note that the local maximum in the $T=104$~mK $R_{yy}$ isotherm
measured near $\nu_c^{R7a}=7.30$, shown in Fig.2a,
may still be associated with the bubble phase $R7a$; this local maximum 
indicates a precursor of the bubble phase $R7a$\cite{deng3}.

\begin{figure}[t]
  \centering
  \includegraphics[width=3.4in]{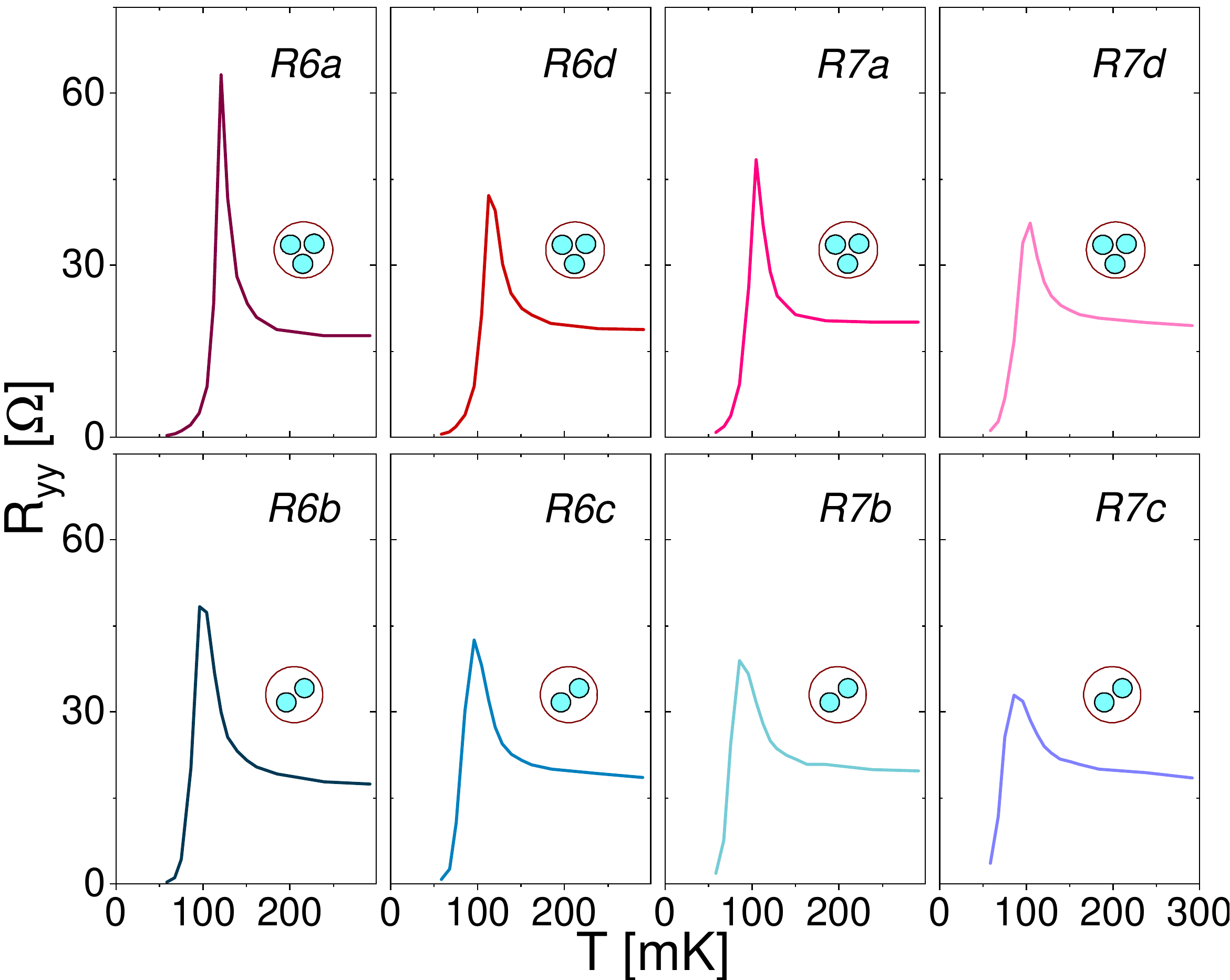}
  \caption{Temperature dependence of the longitudinal magnetoresistance measured at fixed filling factor $\nu=\nu_C$
  for the eight bubble phases in the $N=3$ Landau level. Curves exhibit a sharp peak near the onset
  temperatures $T_{c}$ of the bubble phases.}
\end{figure}

In Fig.2b we plot the evolution of $R_{yy}$ with $T$ as measured at the
central filling factor $\nu_c$ for the bubble phases $R7a$ and $R7b$. We denote such curves as
$R_{yy}(T)|_{\nu=\nu_c}$. These $R_{yy}(T)|_{\nu=\nu_c}$ curves may be thought of as cuts 
along a constant filling factor $\nu=\nu_c$ in the $R_{yy}(\nu,T)$ 
manifold having two independent variables $\nu$ and $T$.
As expected, $R_{yy}(T)|_{\nu=\nu_c}$ is vanishingly small at the lowest measured temperatures,
indicating well developed bubble phases.
In addition $R_{yy}(T)|_{\nu=\nu_c}$ has a finite and nearly $T$-independent value at $T>200$~mK.
However, near $T=T_{c}^{R7a}=101$~mK, $R_{yy}(T)|_{\nu=\nu_c}$ for the $R7a$ phase exhibits a sharp peak.
Similar sharp peaks in $R_{yy}(T)|_{\nu=\nu_c}$ curves were measured at the onset temperatures
of bubble phases in the $N=1$ and $N=2$ Landau levels\cite{deng1,deng2,chick-1}.
As seen in Fig.3, we now detect such peaks for all multielectron bubble phases of the $N=3$ Landau level.
We conclude that, in high mobility samples there is a sharp peak present in the
$R_{yy}(T)|_{\nu=\nu_c}$ curves near the onset of multielectron bubble phases,
irrespective of the Landau level they develop in.

Data available for bubble phases in the $N=3$ Landau level in an alloy sample\cite{zud}
offer a chance for comparison. Because of the deliberately introduced Al into the GaAs channel during the
sample growth process, thereby forming a dilute AlGaAs alloy,
the alloy sample in Ref.\cite{zud} had a mobility of 
$\mu = 3.6 \times 10^{6}$ cm$^{2}$/Vs. This number is about a factor of 4 times less than that of our sample. 
Quite interestingly, in the alloy sample bubble phases of the $N=3$ Landau level develop 
at the same filling factors and also in a similar temperature range as those in our work\cite{zud}. 
A consequence of the reduced mobility, which can be seen at temperature much above bubble onset,
is the enhanced longitudinal magnetoresistance of $\approx 80$~$\Omega$ in the alloy sample\cite{zud}, 
compared to $\approx 18$~$\Omega$ in our sample.
Another consequence is the conspicuous absence of the sharp peak in the $R_{yy}(T)|_{\nu=\nu_c}$
curves\cite{zud}. Indeed, as the temperature increases in the sample with added disorder\cite{zud}, 
the longitudinal resistance of the bubble phase increases and saturates past $~135$ mK, without the 
development of a sharp peak.
Transport in bubble phases is currently understood as follows: at $T << T_c$ the bubble phase is pinned by 
the disorder present in the sample, whereas at $T >> T_c$ electrons form a uniform uncorrelated liquid. 
In this interpretation, near $T=T_c$ these two phases compete by forming an interpenetrating network of domains
throughout the bulk of the sample. The presence of a peak in $R_{yy}(T)|_{\nu=\nu_c}$ 
in our high mobility GaAs sample and also in graphene\cite{dean} in a narrow range of temperatures near $T=T_c$
indicates excess scattering due to enhanced thermal fluctuations between the domains of the two competing phases. 
We think that such thermal fluctuations and the associated sharp resistance peak are suppressed in
the alloy sample by the disorder present\cite{zud}.

We now examine the onset temperatures of bubble phases, quantities
related to the corresponding cohesive energies calculated in Hartree-Fock theories\cite{deng1,deng2}. 
We found that onset temperatures of the $M=2$ and $M=3$ bubble phases in the $N=3$
Landau level are close to each other. This property is consistent with the Hartee-Fock 
predictions\cite{fogler,fogler96,moessner,fogler97,foglerBook,cote03,goerbig03,goerbig04,dorsey06}. 
Quantitative comparisons with calculated cohesive energies are, however, 
tenuous. This is partly because cohesive energies are calculated under idealized conditions,
such as no disorder and no Landau level mixing.
Discrepancies of more than two orders of magnitude between the onset temperatures\cite{deng1,deng2} 
and calculated cohesive energies in the $N=1$ and $N=2$ Landau levels\cite{fogler,fogler96,moessner,fogler97,foglerBook,cote03,goerbig03,goerbig04,dorsey06} 
were indeed attributed to these idealized conditions. We found that these discrepancies persist in the $N=3$ Landau 
level\cite{cote03,goerbig04,dorsey06}. 

Nonetheless, comparisons of onset temperatures and cohesive energies provides useful insight
to the nature of electronic interactions. It is well-known that the clustering of electrons into bubbles is promoted by
competing short-range and long-range electronic interactions\cite{fogler,fogler96,moessner,jim-rev,kevin}. 
The long-range interaction is Coulombic in nature, while
the short-range interaction is a softened Coulomb potential. At the root of such a potential softening we find 
overlapping single particle wavefunctions\cite{fogler,fogler96,moessner,jim-rev,kevin} 
and finite layer thickness effects\cite{layer,dorsey06}. 

\begin{figure}[t]
  \centering
  \includegraphics[width=3.4in]{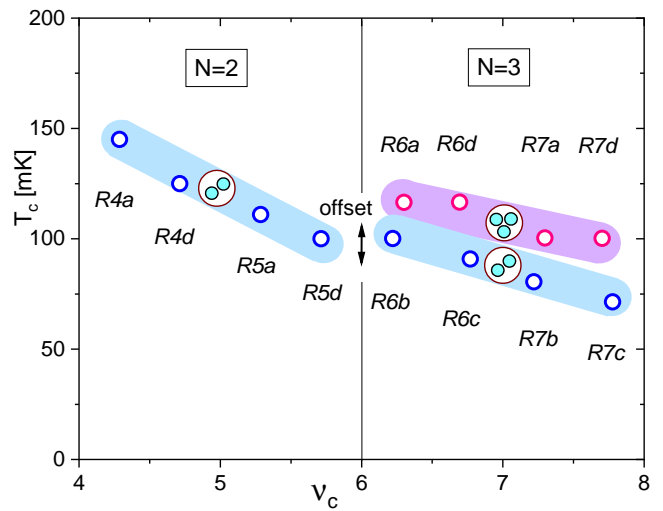}
  \caption{Dependence of the onset temperature $T_c$ on the central factor
  of the $M=2$ and $M=3$ bubble phases in the $N=2$ and $N=3$ Landau levels.
  Shaded bands illustrate trends of onset temperatures for phases with the same number of electrons per bubble.
  Near $\nu=6$, the dimensionless onset temperatures of the $M=2$ bubble phases exhibit an offset marked
  by the double arrow.
  }
\end{figure}

At first sight, the onset temperatures in the $N=3$ Landau level listed in Table.I.
do not seem to follow a particular trend. However, a closer inspection reveals some interesting properties.
Within one Landau level, onset temperatures of a given type of bubble phase form
an approximately linear trend. 
In Fig.4 we show onset temperatures $T_c$ for multielectron bubble phases in
the $N=2$ and $N=3$ Landau levels. The three colored bands in Fig.4 indicate these linear trends for
the $M=2$ bubble phases of the $N=2$ Landau level, for
the $M=2$ bubble phases of the $N=3$ Landau level, and for
the $M=3$ bubble phases of the $N=3$ Landau level.
Since data for bubble phases forming in different spin branches of a given orbital
Landau level lie on the same line, we conclude that onset temperatures are not influenced by the spin quantum number.

Identifying the dominant bubble phase in the $N=3$ Landau level reveals details 
on the short-range electron-electron interaction that drive bubble formation. We note, that 
Hartree-Fock calculations do not provide consistent results for the dominant,
i.e. the most stable, bubble phase. Indeed, in the $N=3$ Landau level 
Refs.\cite{fogler97,dorsey06} predict the $M=3$ bubbles to be dominant, whereas Refs.\cite{cote03,goerbig04} 
find the $M=2$ bubbles to be stronger. The former results agree, but the latter ones are contrary to our findings.
A likely cause of different dominant bubble phase may be different effective electron-electron interaction.
To see this, the work of Ettouhami et al.\cite{dorsey06} is particularly useful. In this work authors tuned the
short-range part of the electron-electron interaction through the layer thickness parameter
$\lambda$, while keeping the long-range Coulombic
potential unchanged\cite{dorsey06}. It was found that in the $N=3$ Landau level the energy balance
can be significantly tilted: the $M=3$ bubbles are dominant
for $\lambda=0$, whereas the $M=3$ bubbles have nearly the same energy with $M=2$ bubbles at 
$\lambda=1$, i.e. when the electron-electron interaction was softened at short distances\cite{dorsey06}. 
We then surmise that a further softening of the potential may reverse the energy balance
of the $M=3$ and $M=2$ bubble phases and therefore may yield the experimentally 
observed dominant bubble phase.

\begin{figure}[t]
  \centering
  \includegraphics[width=3.4in]{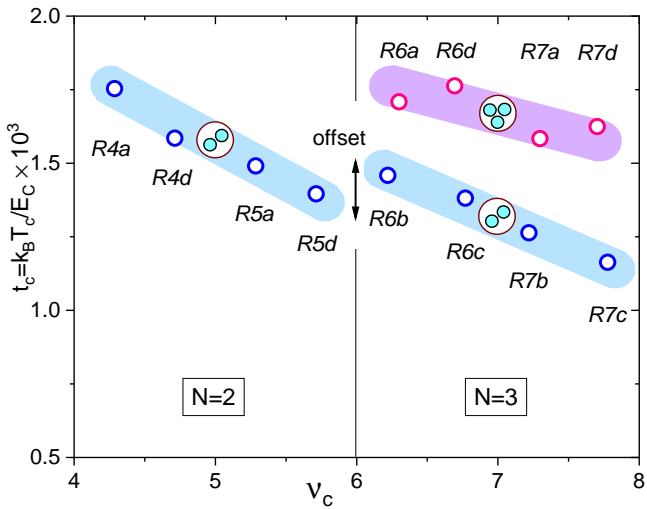}
  \caption{Dependence of the dimensionless onset temperature $t_c=k_B T_c/E_C$ on the central factor
  of the $M=2$ and $M=3$ bubble phases in the $N=2$ and $N=3$ Landau levels.
  Shaded bands illustrate trends of onset temperatures for phases with the same number of electrons per bubble.
  Near $\nu=6$, the dimensionless onset temperatures of the $M=2$ bubble phases exhibit an offset marked
  by the double arrow.
  }
\end{figure}

A comparison of the energetics of $M=2$ bubble phases in the $N=2$ and 
$N=3$ Landau levels reveals that the electronic short-range interaction is dependent on the
Landau index $N$. We discussed earlier the linear trend of both $T_c$ 
versus $\nu_c$ for the $M=2$ bubble phases. These linear trends
exhibit a vertical offset when $N$ changes from $2$ to $3$ in the vicinity of $\nu=6$. 
Indeed, as seen in Fig.4, the two colored bands associated with $M=2$ bubble phases
in the $N=3$ Landau level acquired an offset when compared to that for $M=2$ bubble phase
in the $N=2$ Landau level. We attribute this offset to a variation of the
effective electron-electron interaction, specifically its short-range part, with the Landau index $N$. 
While finite layer thickness effects soften the electron-electron interaction, they are not expected
to depend on the Landau level index. In contrast, a short-range potential that results from the
overlapping single electron wavefunctions is Landau index dependent\cite{fogler,fogler96,moessner,jim-rev,kevin}. 
This is because the number of nodes in these wavefunctions directly influences bubble energetics.
The comparison of the energetics of the $M=2$ bubble phases in the $N=2$ and $N=3$
Landau levels thus provides direct evidence that the overlapping electronic wavefunctions
play a role in shaping the short-range part of the electron-electron interaction.

In the following, we examine properties of the dimensionless onset temperatures $t_c=k_B T_c/E_C$,
a quantity closely related to the dimensionless cohesive energy of Hartree-Fock calculations.
Here $k_B$ is the Boltzmann constant, $E_C=e^2/4 \pi \epsilon l_B$
the Coulomb energy, and $l_B$ is the magnetic length. 
Dimensionless onset temperatures $t_c$
for the multielectron bubble phases in the $N=2$ and $N=3$ Landau levels are plotted in Fig.5.
Trends already discussed for $T_c$ are also observed for $t_c$: within one Landau level, 
both of these quantities have a linear trend with $\nu_c$ and these  linear trends
exhibit a vertical offset when $N$ changes from $2$ to $3$ in the vicinity of $\nu=6$. 
In addition, we find that across the different Landau levels, the linear trend of $t_c$ versus $\nu_c$ for
the $M=2$ bubble phases have a similar slope, $\partial t_c/ \partial \nu_c \approx - 2.5 \times 10^{-4}$
in both the $N=2$ and $N=3$ Landau levels. We thus found that bubble phases with the
same number of electrons forming in different Landau levels share a similar $\partial t_c/ \partial \nu_c$.
In contrast, the $M=3$ bubble phases in the $N=3$ Landau level have a significantly diminished
$\partial t_c/ \partial \nu_c$ slope, reduced by about a factor $5$ as compared to that of the $M=2$ bubble phases. 

In conclusion, we observed qualitative and quantitative aspects of bubble formation in the $N=2$
and $N=3$ Landau levels. We found that in our high mobility sample the longitudinal magnetoresistance 
versus temperature curves exhibit sharp peaks in the multielectron bubble regions both in the 
$N=2$ and $N=3$ Landau levels. We used these peaks to extract the onset temperatures for the bubble phases.
The recent assignment of the number of electrons per bubble to these phases
allowed an analysis of the measured onset temperatures. We found that within the $N=3$ Landau level,
onset temperatures of different bubble phases exhibit linear trands with the filling factor.
However, the onset temperatures of the $M=3$ bubbles are higher than those of $M=2$ bubbles.
Furthermore, when comparing the $M=2$ bubble phases across $N=2$ and $N=3$ Landau levels
we found that they are similar, but they exhibit an offset.
These measurements offer information on bubble energetics that is expected to lead to refinements
of existing theories and offer evidence that short-range electron-electron 
interactions originating from overlapping wavefunctions are at play in bubble formation.

Measurements at Purdue were supported by the NSF grant  DMR 1904497. The sample growth effort of M.J.M.
was supported by the DOE BES award DE-SC0006671, while that of L.N.P. and K.W.W. by the Gordon
and Betty Moore Foundation Grant No. GBMF 4420, and the NSF MRSEC Grant No. DMR-1420541.

\end{document}